\documentclass[conference]{IEEEtran}
\IEEEoverridecommandlockouts
\usepackage{cite}
\usepackage{amsmath,amssymb,amsfonts}
\usepackage{algorithmic}
\usepackage{mathtools}
\usepackage{graphicx}
\usepackage{textcomp}
\usepackage{xcolor}

\usepackage[pdftex]{hyperref}
\usepackage{tikz}
\newcommand\copyrighttext{%
    \footnotesize \textcopyright 2025 IEEE. Personal use of this material is permitted.
    Permission from IEEE must be obtained for all other uses, in any current or future
    media, including reprinting/republishing this material for advertising or promotional
    purposes, creating new collective works, for resale or redistribution to servers or
    lists, or reuse of any copyrighted component of this work in other works.
    DOI: \href{https://doi.org/10.1109/BigData66926.2025.11401461}{https://doi.org/10.1109/BigData66926.2025.11401461}}
\newcommand\copyrightnotice{%
    \begin{tikzpicture}[remember picture,overlay]
        \node[anchor=south,yshift=10pt] at (current page.south) {\fbox{\parbox{\dimexpr\textwidth-\fboxsep-\fboxrule\relax}{\copyrighttext}}};
    \end{tikzpicture}%
}

\hypersetup{pdfcreator={}, pdfproducer={}}
    
\begin{document}

\title{Learning Process Energy Profiles from  \\ Node-Level Power Data\\}

\author{
    \IEEEauthorblockN{Jonathan Bader\IEEEauthorrefmark{1}, Julius Irion\IEEEauthorrefmark{1}, Jannis Kappel\IEEEauthorrefmark{1}, Joel Witzke\IEEEauthorrefmark{1}, Niklas Fomin, Diellza Sherifi,
        and Odej Kao}

    \IEEEauthorblockA{
        \{firstname.lastname\}@tu-berlin.de, Technische Universität Berlin, Germany\\
    }

}
\IEEEpubid{\makebox[\columnwidth]{*equal contribution, alphabetical order \hfill} \hspace{\columnsep}\makebox[\columnwidth]{ }}

\maketitle
\copyrightnotice

\begin{abstract}
The growing demand for data center capacity, driven by the growth of high-performance computing, cloud computing, and especially artificial intelligence, has led to a sharp increase in data center energy consumption.
To improve energy efficiency, gaining process-level insights into energy consumption is essential.
While node-level energy consumption data can be directly measured with hardware such as power meters, existing mechanisms for estimating per-process energy usage, such as Intel RAPL, are limited to specific hardware and provide only coarse-grained, domain-level measurements.

Our proposed approach models per-process energy profiles by leveraging fine-grained process-level resource metrics collected via eBPF and perf, which are synchronized with node-level energy measurements obtained from an attached power distribution unit.

By statistically learning the relationship between process-level resource usage and node-level energy consumption through a regression-based model, our approach enables more fine-grained per-process energy predictions.

\end{abstract}



\section{Introduction} 
\label{sec:introduction}
Over the past decade, data center energy consumption growth has accelerated significantly. 
Advancements in artificial intelligence are driving increased demands on data centers, especially with the rise of Generative AI models, which require immense processing power to train and run such models~\cite{bashir2024climate, wu2024beyond}. 
Today, data centers make up around 1.5\% of the global electricity consumption, and their share is expected to more than double by 2030~\cite{iea2025energyai}. 
The growing energy demands driven by AI have resulted in overloads and capacity constraints across local grids that supply large-scale data centers~\cite{lin2024exploding}.
To maintain grid energy availability, coal and nuclear power plants may be reopened or newly constructed, potentially increasing environmental strain~\cite{noland2024will, kyriakarakos2025artificial}.

With growing concerns about carbon emissions and rising energy costs~\cite{iea2025energyai}, improving energy efficiency can relieve the need for expensive infrastructure overhauls or increased energy supply~\cite{huang2020review, Katal2023energy}. 
To improve practical efficiency, ML workflows could be adapted, software optimized, or entire pipelines assigned to different machines based on energy availability and load patterns~\cite{hussain2022deadline, wiesner2021let}.
However, to make these decisions, detailed process-level insights into energy consumption are necessary.

Currently, hardware monitoring tools such as external power meters and software tools such as \textit{nvidia-smi}, \textit{powertop}, or \textit{powerstat}, are widely available and can be used to track energy usage on a per-node basis.
They provide insights into the combined energy usage of all components and processes running on a specific node. 
However, energy usage is only measured on a per-node basis, while a nodes usually runs multiple processes in parallel.
The energy usage of a single process can therefore not easily be estimated \cite{guan2023wattscope, leonvega2025efimon}.
Existing approaches rely on hardware-specific tools such as Intel's Running Average Power Limit (RAPL)~\cite{jay2023experimental}. 
While it does provide fine-grained power usage estimates, RAPL support is hardware-specific, full domain coverage and kernel-level support is largely constrained to Intel CPUs, whereas AMD implementations expose only partial interfaces~\cite{raffin2024dissecting}. 
This does not make it suitable for many data center operators~\cite{khan2018}. 
Therefore, a hardware-agnostic approach to per-process energy usage estimation is needed.

\begin{figure}[t!]
	\centering
	\includegraphics[width=1\columnwidth]{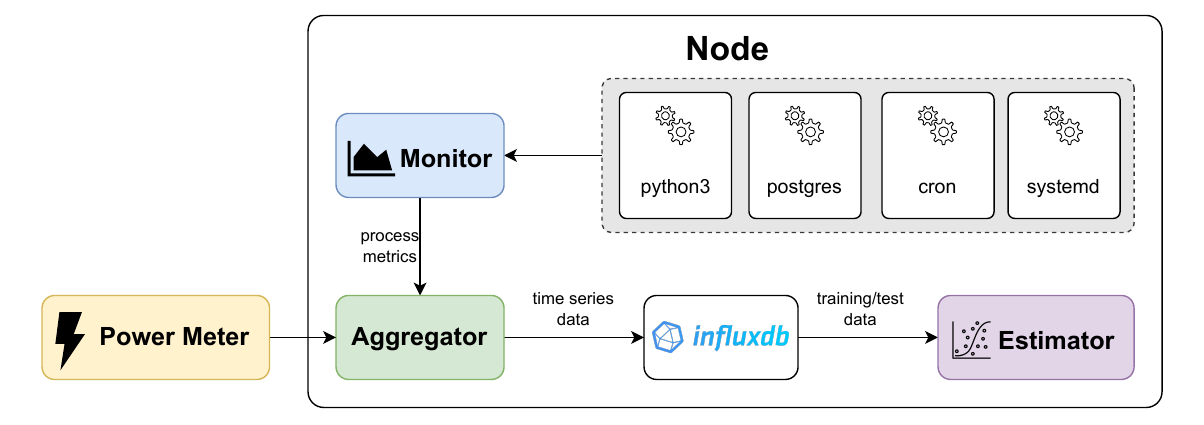}
	\caption{Overall system architecture and the interaction between monitoring component, aggregator component, and estimator component.}
	\label{fig:system-architecture}
\end{figure}

Figure~\ref{fig:system-architecture} presents our system that learns per-process energy profiles by correlating process-level resource usage metrics with node-level energy consumption data using a regression model.
We leverage fine-grained monitoring tools such as eBPF and perf to collect detailed performance metrics.
By statistically modeling the relationship between these metrics and the energy usage from smart meters, our system provides insights into the power usage of individual processes.
We provide a publicly available prototype implementation of our system\footnote{github.com/dos-group/energy-process-mapping} and evaluate it using real-world processes on our commodity cluster.

\section{Approach}
\label{sec:approach}

In this section, we present our approach for estimating per-process energy consumption from node-level power data.
Figure~\ref{fig:system-architecture} gives a high-level overview of the involved components.

\subsection{Energy and Resource Monitoring}
\label{subsec:data_collection}

Our energy and resource monitoring component, visible in Figure\ref{fig:system-architecture}, is designed to provide a holistic view of system and process-level resource usage, focusing on capturing detailed metrics that are crucial for modeling energy consumption at the granularity of individual processes. 
We integrate multiple sources of data: system-level metrics obtained from kernel-level monitoring using \textit{eBPF}, process-specific metrics obtained through tools like \texttt{psutil}, and node-level power measurements from an external smart meter.

Resource monitoring is achieved by the continuous collection of process-level metrics using kernel-level instrumentation through our eBPF monitor client, which collects data such as CPU time, memory usage, disk I/O, network traffic, and context switches for each active process in the system. 
Additionally, process-specific data is enriched by integrating information from the linux \texttt{psutil}-interface, which provides an alternative source for detailed accumulated CPU time measurements since process start.

Energy consumption data is captured by periodically querying the smart meter which provides a REST interface and directly measures the power consumed by the entire system. 
To capture short-term fluctuations, power readings are sampled multiple times within each monitoring interval and then averaged to obtain a single representative value, thereby increasing the temporal resolution of the data. 
The data is aggregated over fixed intervals, with adjustable sampling rates.

During each 1 s interval, node-level power is sampled at 4 Hz (every 250 ms) and averaged over the window, while per-process resource metrics are captured once per interval and converted into deltas. 
The aggregator component computes deltas between process-level metrics of consecutive intervals to represent the actual activity within each interval, since many system counters (e.g., CPU time) are cumulative by nature.

\subsection{Regression-Based Energy Modeling}
\label{subsec:energy_modeling}

Our regression-based model estimates the energy consumption of processes by learning from our monitored data. 
Rather than assuming a direct proportionality between CPU usage and energy consumption, the model captures complex relationships between various system features, including CPU utilization, memory consumption, disk I/O, network activity, and other system parameters that may influence energy consumption.

The prediction model includes a static component representing the baseline power, which is added to all per-process estimations.
The sum of these contributions is compared to the total node energy consumption measured by the smart meter. Let \(y_t\) denote the observed total energy in interval \(t\). Each process \(r\) has a standardized feature vector \(\tilde x_r\in\mathbb{R}^p\) assigned per interval. The assignment is encoded by \(A\in\{0,1\}^{T\times N}\) with \(A_{tr}=1\) iff \(r\) belongs to interval \(t\).
\begin{align}
z_t &\coloneqq \sum_{r=1}^{N} A_{tr}\,\tilde x_r, \qquad t=1,\dots,T.
\end{align}
The implementation estimates a vector of nonparametric marginal costs \(w \in \mathbb{R}^p\) and a static component \(s \in \mathbb{R}_{\ge 0}\) by solving a convex, \(\ell_1\)-regularized least-squares problem for with an explicit non-negativity constraint on \(s\):
\begin{align}
\min_{w,\, s \ge 0}\;\;
& \sum_{t=1}^{T} \big(y_t -(z_t^{\top} w + s\big))^{2}
\;+\; \lambda_1 \,\lVert w\rVert_1 \;+\; \lambda_2 |s|.
\end{align}
The first term penalizes interval-level residuals, \(\lambda_1\) promotes sparsity for identifiability under aggregation and noise, and \(\lambda_2\) regularizes the static baseline term to minimize it. The problem is modeled and solved to global optimality using CVXPY, yielding the weights \(\hat w\) and the static component \(\hat s\).

\section{Evaluation}
\label{sec:eval}
In this section, we describe the evaluation of our per-process energy approach for accuracy, robustness, and practicality.

\subsection{Experimental Setup}
In our experiments, we created a load generation script using the Phoronix Test Suite (PTS) in conjunction with a fixed set of benchmarks from Openbenchmarking to create realistic workloads.
The benchmarks used were selected based on their relevance for real-world data center scenarios. 
The load script orchestrated phases of compute-intensive activity interleaved with idle periods. 

For regression training, the dataset was split in a time-aware manner into a contiguous 80/20 train–test partition, ensuring that later intervals served as the held-out test set.

Power measurements were obtained via the API of the GUDE Expert Power Control 8045 smart meter, which provides per-outlet sampling of active power.
The experiments were conducted on a dedicated node running Ubuntu 22.04, equipped with an AMD EPYC 8224P, 192 GB RAM, and two 4 TB NVMe SSDs. 

\begin{figure}[t]
    \centering
    \includegraphics[width=1.0\linewidth]{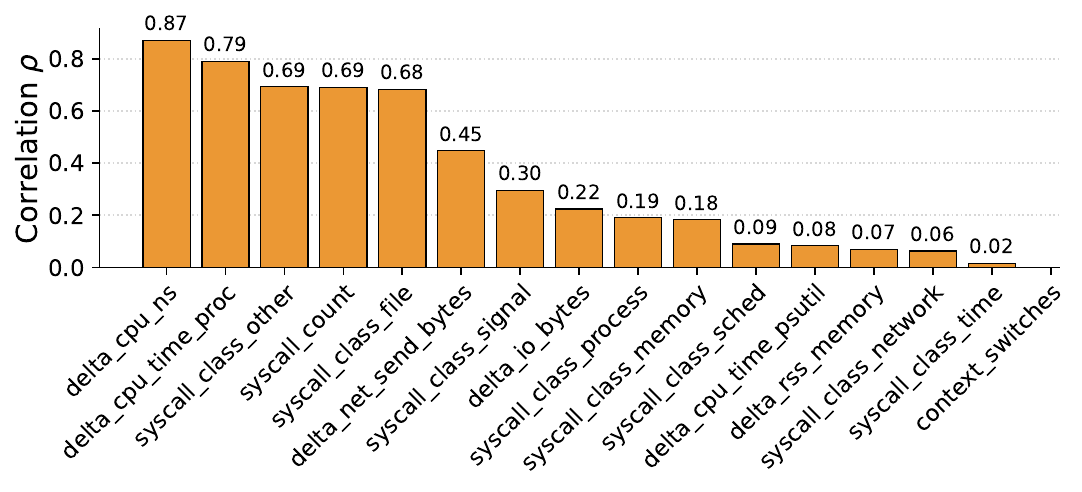}
    \caption{Spearman correlation of monitored features with interval energy}
    \label{fig:spearman_correlation}
\end{figure}

\subsection{Correlation Analysis}

We conducted a Spearman correlation analysis to identify the importance of monitored resources on the aligned interval energy consumption.
Figure~\ref{fig:spearman_correlation} shows that CPU-centric metrics from independent accounting sources exhibited the strongest associations with interval energy, with $\rho$ in the range of 0.8--0.9. 
Selected syscall and network features display moderate dependence.
Memory and context-switch counts are weakly associated or yield a $\rho$ near zero.

These analysis motivated a compact feature set centered on CPU time and a few complementary kernel indicators. 
We excluded network-related metrics due to noise from SSH activity. 
This reduced set improves interpretability while retaining sufficient predictive accuracy by limiting collinearity and reducing variance. The model therefore yields physically meaningful energy weights that generalize across workloads while remaining lightweight and hardware-agnostic.

\subsection{Experimental Results}

\begin{figure}[t]
  \centering
  \includegraphics[width=1.0\linewidth]{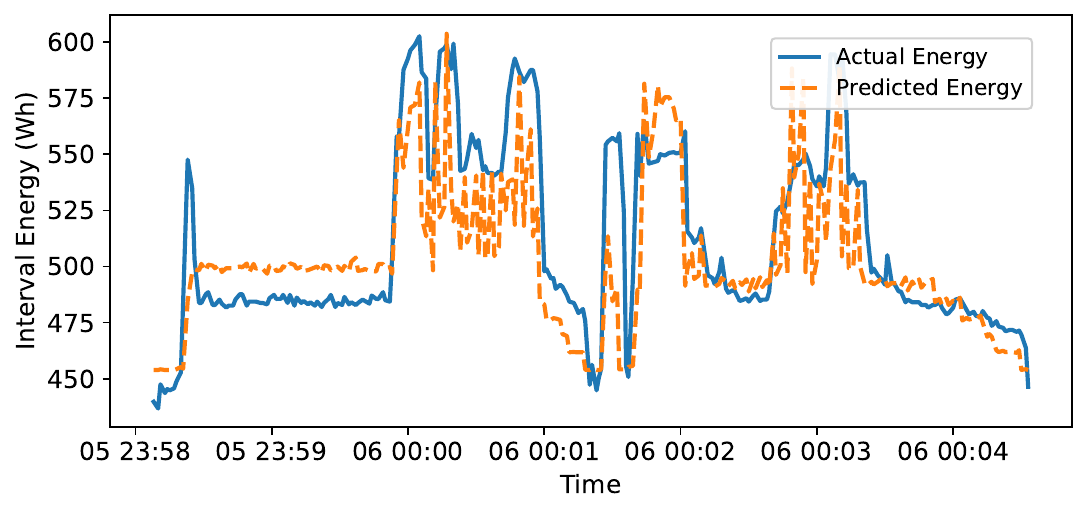}
  \caption{Estimated overall energy versus real overall energy}
  \label{fig:comparison_regression_real}
\end{figure}

A comparison of the actual energy usage against our predicted energy usage is shown in Figure~\ref{fig:comparison_regression_real}. 
The results show a $R^2 \approx 0.59$ with $MAE \approx 17.67$, which corresponds to approximately 3.5\% of the average measured energy. Over time and during different phases such as compute, idle, and transition, the predicted energy closely follows the measured consumption.
The main differences stem from sharp load transitions and unmodeled power-management effects such as dynamic voltage and frequency scaling (DVFS) or core idling. 
Overall, the regression reliably reproduces the temporal evolution of the node’s energy usage, reflecting the stability and explanatory strength of the selected metrics. 

\begin{figure}[t]
  \centering
  \includegraphics[width=1.0\linewidth]{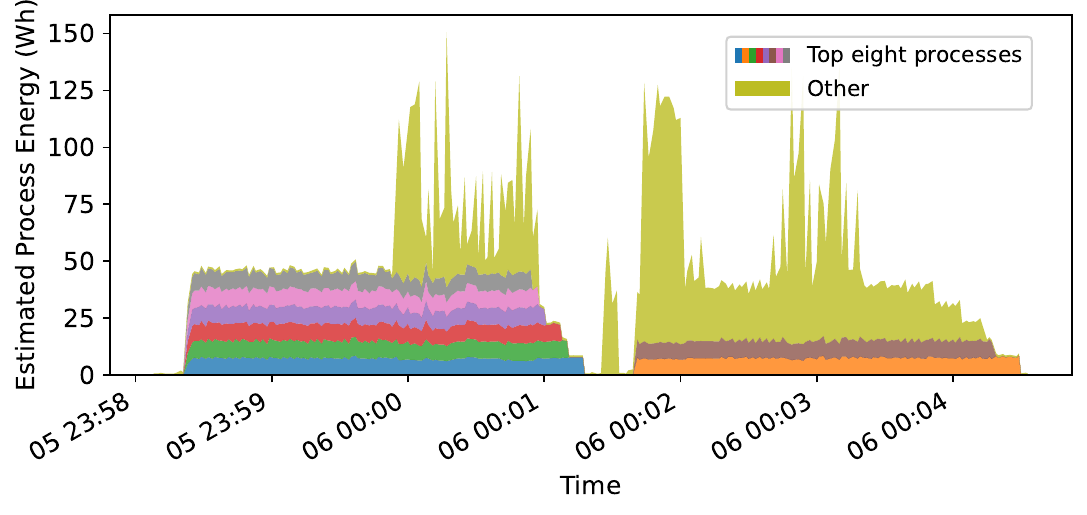}
  \caption{Per-process energy estimated by the model}
  \label{fig:decomposed_regression}
\end{figure}

Figure~\ref{fig:decomposed_regression} illustrates the estimated per-process energy contributions across time, highlighting the top eight processes alongside an aggregated “Other” category. 
During the initial compute-intensive phase, energy usage made up of several contributing processes remains relatively stable. 
When a new energy-consuming process appears, the estimated contributions of other processes remain largely unaffected, indicating that the model successfully attributes energy independently to the correct sources. 
As the system transitions to idle phases, overall energy usage decreases accordingly, while relative process proportions are preserved.


\section{Conclusion}
\label{sec:conc}
This paper presented an approach to predict process-level energy consumption by learning a regression model that combines low-level resource monitoring with coarse-grained node energy measurements.
The evaluation of our open-source implementation showed a prediction error of about 3.5\% percent of the mean interval energy, demonstrating the potential for accurate modeling of process energy consumption.

\section*{Acknowledgments}
\thanks{Funded by the Deutsche Forschungsgemeinschaft (DFG, German Research Foundation) as FONDA (Project 414984028, SFB 1404).}

\bibliographystyle{IEEEtran}
\bibliography{bibliography}

\end{document}